\documentclass[runningheads]{llncs}
\usepackage{graphicx}
\usepackage{amsmath}
\usepackage{amsfonts}
\usepackage[colorlinks]{hyperref}
\hypersetup{
    colorlinks=true,
    linkcolor=blue,
    filecolor=blue,      
    urlcolor=blue,
    citecolor=blue,
}

\begin{document}
\title{Learned Proximal Networks for Quantitative Susceptibility Mapping}
\titlerunning{Learned Proximal Networks for QSM}
%
\author{Kuo-Wei Lai\inst{1,2}
\and
Manisha Aggarwal\inst{3}
\and
Peter van Zijl\inst{2,3}
\and
Xu Li\inst{2,3}
\and
Jeremias Sulam\inst{1}
}
\authorrunning{Lai et al.}

\institute{Department of Biomedical Engineering, Johns Hopkins University, Baltimore, MD 21218, USA\\
\email{klai10@jhu.edu}
\and
F.M. Kirby Research Center for Functional Brain Imaging, Kennedy Krieger Institute, Baltimore, MD 21205, USA\\
\and
Department of Radiology and Radiological Sciences, Johns Hopkins University, Baltimore, MD 21205, USA\\
}

\maketitle              
%
\begin{abstract}
Quantitative Susceptibility Mapping (QSM) estimates tissue magnetic susceptibility distributions from Magnetic Resonance (MR) phase measurements by solving an ill-posed dipole inversion problem. Conventional single orientation QSM methods usually employ regularization strategies to stabilize such inversion, but may suffer from streaking artifacts or over-smoothing. Multiple orientation QSM such as calculation of susceptibility through multiple orientation sampling (COSMOS) can give well-conditioned inversion and an artifact free solution but has expensive acquisition costs. On the other hand, Convolutional Neural Networks (CNN) show great potential for medical image reconstruction, albeit often with limited interpretability. Here, we present a Learned Proximal Convolutional Neural Network (LP-CNN) for solving the ill-posed QSM dipole inversion problem in an iterative proximal gradient descent fashion. This approach combines the strengths of data-driven restoration priors and the clear interpretability of iterative solvers that can take into account the physical model of dipole convolution. During training, our LP-CNN learns an implicit regularizer via its proximal, enabling the decoupling between the forward operator and the data-driven parameters in the reconstruction algorithm. More importantly, this framework is believed to be the first deep learning QSM approach that can naturally handle an arbitrary number of phase input measurements without the need for any ad-hoc rotation or re-training. We demonstrate that the LP-CNN provides state-of-the-art reconstruction results compared to both traditional and deep learning methods while allowing for more flexibility in the reconstruction process. 

\end{abstract}
%

\section{Introduction}
Quantitative Susceptibility Mapping (QSM) is a Magnetic Resonance Imaging (MRI) technique that aims at mapping tissue magnetic susceptibility from gradient echo imaging phase \cite{de2008quantitative,wang2015quantitative}. QSM has important clinical relevance since bulk tissue magnetic susceptibility provides important information about tissue composition and microstructure \cite{liu2015susceptibility,wang2015quantitative} such as myelin content in white matter and iron deposition in gray matter. Pathological changes in these tissue susceptibility sources are closely related to a series of neurodegenerative diseases such as Multiple Sclerosis \cite{chen2014quantitative,langkammer2013quantitative,li2016magnetic} and Alzheimer's Disease \cite{acosta2013vivo,ayton2017cerebral,van2016colocalization}.

The QSM processing typically includes two main steps \cite{deistung2017overview}: (i) a phase preprocessing step comprising phase unwrapping and background field removal, which provides the local phase image, and (ii) the more challenging phase-to-susceptibility dipole inversion, which reconstructs the susceptibility map from the local phase. Due to the singularity in the dipole kernel and the limited number of phase measurements in the field of view, inverting the dipole kernel is an ill-posed inverse problem \cite{deistung2017overview}. Different strategies have been used in conventional QSM methods to stabilize this ill-posed inverse. One strategy is through data over-sampling, i.e. by utilizing multiple phase measurements acquired at different head-orientations to eliminate the singularity in the dipole kernel. Such calculation of susceptibility through multiple orientation sampling (COSMOS) is usually used as a gold standard for in vivo QSM \cite{liu2009calculation}. Even though COSMOS QSM exhibits excellent image quality, it is often prohibitively expensive in terms of data acquisition, as phase images at three or more head orientations are required, leading to long scan times and patient discomfort. For single orientation QSM, direct manipulation of the dipole kernel as in thresholded k-space division (TKD) often leads to residual streaking artifacts and susceptibility underestimation \cite{shmueli2009magnetic}. To suppress such streaking artifacts \cite{deistung2017overview,langkammer2018quantitative}, different regularization strategies are often used, e.g. as in morphology enabled dipole inversion (MEDI) \cite{liu2011morphology,liu2013nonlinear} and in the two-level susceptibility estimation method, STAR-QSM \cite{wei2015streaking}. However, such methods usually need careful parameter tuning or may introduce extra smoothing due to the regularization enforced, leading to sub-optimal reconstruction.

Upon the advent of deep convolutional neural networks in a myriad of computer vision problems, some recent deep learning based QSM algorithms have shown that these data-driven methods can approximate the dipole inversion and generate high-quality QSM reconstructions from single phase measurements \cite{jung2020overview}. QSMnet \cite{yoon2018quantitative} used a 3D Unet \cite{ronneberger2015u} to learn a single orientation dipole inversion and its successor, QSMnet$^{+}$ \cite{jung2020exploring}, explored model generalization on wider susceptibility value ranges with data augmentation. DeepQSM \cite{bollmann2019deepqsm} showed that a deep network trained with synthetic phantom data can be deployed for QSM reconstruction with real phase measurement. More recently, Chen et al. \cite{chen2020qsmgan} utilized generative adversarial networks \cite{goodfellow2014generative} to perform dipole inversion, while Polak et al. \cite{NDI} proposed a new nonlinear dipole inversion and combined it with a variational network (VaNDI) \cite{hammernik2018learning} to further improve accuracy. Closest to our work is that from Kames et al. \cite{pvn}, who employed a proximal gradient algorithm with a variational network. However their approach is limited to downsampled data due to GPU memory constraints, limiting its applicability.

While these previous works illustrate the potential benefit of deep learning for QSM, they have known limitations \cite{jung2020overview}: most of them \cite{bollmann2019deepqsm,chen2020qsmgan,jung2020exploring,yoon2018quantitative} only learn an unique dipole inversion for a predefined direction (often aligned with the main field), thus having to rotate the input image when it comes to different oblique acquisitions before applying their reconstruction. More importantly, these methods can only take one phase measurement at deployment, preventing further improvements when another phase image is available. They are also quite limited in combining multiple phase measurements for estimation of more complicated models such as used in susceptibility tensor imaging (STI) \cite{li2017susceptibility,liu2010susceptibility}. Inspired by \cite{adler2018learned,mardani2018deep}, we present a Learned Proximal Convolutional Neural Network (LP-CNN) QSM method that provides far more flexibility for solving the ill-posed dipole inversion problem and high quality reconstruction.\footnote{We have recently become aware of a similar work being independently and concurrently proposed in \cite{proxvnet}. That work did not consider employing different input phases, and differ in several implementation details.} We devised an unrolled iterative model combining a proximal gradient descent algorithm and a convolutional neural network. In this way, we naturally decouple the utilization of the forward model and the data-driven parameters within the reconstruction algorithm. As a result, our LP-CNN can handle a single or an arbitrary number of phase input measurements without any ad-hoc rotation or re-training, while achieving state-of-the-art reconstruction results.

\section{Methods}
\begin{figure}[t]
\begin{centering}
\includegraphics[width=\textwidth]{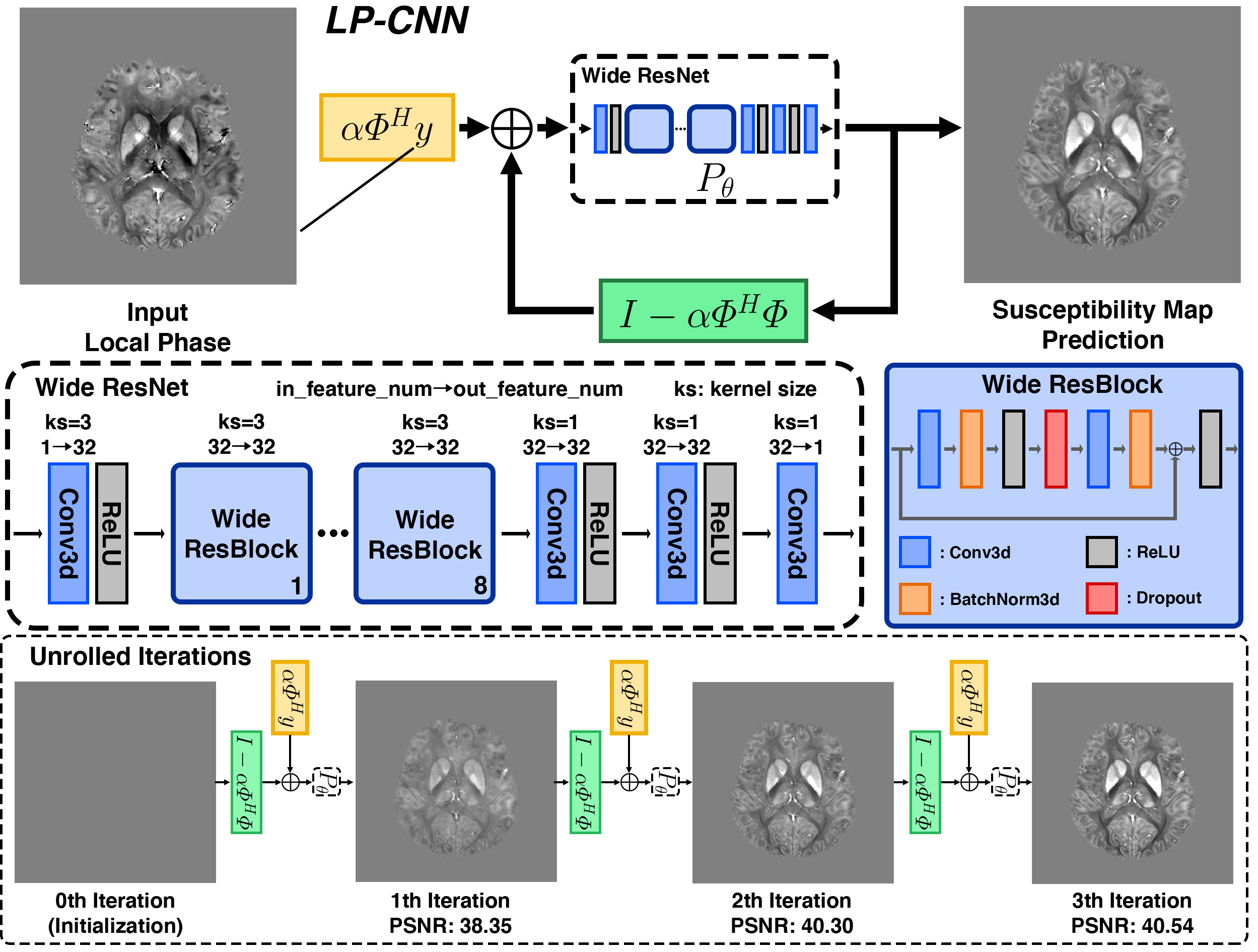}
\caption{A scheme of LP-CNN model (top), Wide ResNet architecture (middle) and unrolled iterations (bottom).} \label{fig1}
\end{centering}
\end{figure}

Our LP-CNN architecture is shown in Fig.~\ref{fig1}. We first preprocess the dataset and then conduct the training of our LP-CNN for dipole inversion. Training data pairs include the local phase data as input data, the corresponding dipole kernel as forward operator, and the COSMOS susceptibility maps as ground-truth data. In order to learn the prior information between local phase measurements and susceptibility maps, we train our LP-CNN in a supervised manner. In the rest of this section, we provide further details for each step.

\subsubsection{Dataset Acquisition and Preprocessing} A total of 36 MR phase measurements were acquired at 7T (Philips Achieva, 32 channel head coil) using three slightly different 3D gradient echo (GRE) sequences with voxel size of \mbox{$1\times1\times1$ mm$^3$} on 8 healthy subjects (5 orientations for 4 subjects with TR=28 ms, TE1/$\delta$TE=5/5 ms, 5 echoes, FOV=$224\times224\times126$ mm$^3$, 4 orientations for 3 subjects with TR=45 ms, TE1/$\delta$TE=2/2 ms, 9 echoes, FOV=$224\times224\times110$ mm$^3$, 4 orientations for 1 subject with TR=45 ms, TE1/$\delta$TE=2/2 ms, 16 echoes, FOV=$224\times224\times110$ mm$^3$) at F.M. Kirby Research Center for Functional Brain Imaging, Kennedy Krieger Institute. The study was IRB approved and signed informed consent was obtained from all participants. GRE magnitude data acquired on each subject at different orientations were first coregistered to the natural supine position using FSL FLIRT \cite{jenkinson2001global}, after which the dipole kernel corresponding to each head orientation was calculated based on the acquisition rotation parameters and coregistration transformation matrix. Multiple steps of phase preprocessing were conducted in the native space of each head orientation including best-path based phase unwrapping \cite{abdul2005fast}, brain masking with FSL BET \cite{smith2002fast}, V-SHARP \cite{ozbay2017comprehensive} for removing the background field and echo averaging for echoes with TEs between 10 ms and 30 ms. The co-registration transformation matrices were then applied on the corresponding local field maps to transform them to the natural supine position. The COSMOS reconstructions were then generated using all 4 or 5 orientations for each subject. Following \cite{yoon2018quantitative}, we also augmented our dataset by generating, in total, 36 simulated local phase maps at random head orientations (within $\pm45^{\circ}$ of the z-axis) using the corresponding COSMOS map as the susceptibility source.

\subsubsection{Proximal Gradient Descent QSM} The dipole inversion problem in QSM can be regarded as an ill-posed linear inverse problem. We model the measurement process by $y = \Phi x + v$,
where $y\in\mathcal{Y}\subseteq\mathbb{R}^n$ is the local phase measurement, $x\in\mathcal{X}\subseteq \mathbb{R}^n$ is the underlying susceptibility map, and $\mathcal{Y}$  and $\mathcal{X}$ are the spaces of phase measurements and susceptibility maps, respectively. Furthermore, $\Phi:\mathcal{X}\to\mathcal{Y}$ is the corresponding forward operator, and $v$ accounts for the measurement noise (as well as accounting for model mismatch). According to the physical model between local phase and susceptibility \cite{liu2015quantitative}, the forward operator is determined by the (diagonal) dipole kernel $D$ as $\Phi = F^{-1}D F$, where $F$ is the (discrete) Fourier transform. Since $D$ contains zeros in its diagonal \cite{liu2015quantitative}, the operator $\Phi$ cannot be directly inverted. In other words, there exist an infinite number of feasible reconstructions $\hat{x}$ that explain the measurements $y$. We can nonetheless formulate this problem as a regularized inverse problem:
\begin{equation} \label{eq:regularized_LS}
\min_x \mathcal{L}(x) = \underbrace{\frac{1}{2}\|y- \Phi x\|^2_2}_{f(x)} + \psi(x),
\end{equation}
where $f$ is a data-likelihood or data-fidelity term and $\psi$ is a regularizer that restricts the set of possible solutions by promoting desirable properties according to prior knowledge\footnote{In fact, the expression in \eqref{eq:regularized_LS} can also be interpreted as a Maximum a Posteriori (MAP) estimator.}. Our goal is to find a minimizer for $\mathcal{L}(x)$ as the estimator for the susceptibility map, $\hat{x}$. Because of its computational efficiency (requiring only matrix-vector multiplications and scalar functions), and ability to handle possibly non-differentiable regularizers $\psi$, we apply The Proximal Gradient Descent algorithm \cite{parikh2014proximal} to minimize \eqref{eq:regularized_LS} iteratively by performing the updates:
\begin{align} \label{eq:proximal} 
\hat{x}_{i+1} &= P_\psi \left(\hat{x}_i - \alpha \nabla f(\hat{x}_i)\right)
\end{align}
where\footnote{$A^H$ denotes the conjugate transpose (or Hermitian transpose) of the matrix $A$.} $\nabla f(\hat{x}) = \Phi^H(\Phi \hat{x}-y)$, $\alpha$ is a suitable gradient step size and $P_\psi(x)$ is the proximal operator of $\psi$, namely $P_{\psi}(x) = \arg\min_u \psi(u) + \frac{1}{2}\|x-u\|^2_2$. Such a proximal gradient descent algorithm is very useful when a regularizer is non-differentiable but still \emph{proximable} in closed form. Clearly, the quality of the estimate $\hat{x}$ depends on the choice of regularizer, $\psi$, with traditional options including sparsity \cite{bruckstein2009sparse} or total variation \cite{osher2005iterative} priors. In this work, we take advantage of supervised learning strategies to parameterize these regularizers \cite{adler2018learned,mardani2018deep,sulam2019multi} by combining proximal gradient descent strategies with a convolutional neural network, as we explain next.

\subsubsection{Learned Proximal CNN} In order to exploit the benefits of deep learning in a supervised regime, we parameterize the proximal operator $P_\psi$ by a 3D Wide ResNet \cite{zagoruyko2016wide} which learns an implicit data-driven regularizer by finding its corresponding proximal operator. This way, Eq. \eqref{eq:proximal} becomes:
\begin{align}\label{eq:learned_prox}
\hat{x}_{i+1} &= P_\theta \left(\hat{x}_i - \alpha \Phi^H(\Phi \hat{x}_i-y)\right) = P_\theta \left( \alpha \Phi ^H y + (I -\alpha \Phi^H \Phi)\hat{x}_i\right),
\end{align}
where $\theta$ represents all learnable parameters in the 3D Wide ResNet. 
The iterates are initialized from $\hat{x}_0=0$ for simplicity\footnote{Any other susceptibility estimation choices can serve as an initialization as well.}, and the final estimate is then given by the $k^{th}$ iteration, $\hat{x}_k$. We succinctly denote the overall function that produces the estimate as  $\hat{x}_k = \phi_{\theta}(y)$. The training of the network is done through empirical risk minimization of the expected loss. In particular, consider training samples $\{y^j,x^j_c\}^{N}_{j=1}$ of phase measurements $y^j$ and target susceptibility reconstructions $x^j_c$. Note that one does not have access to ground-truth susceptibility maps in general, and tissue susceptibility is treated as isotropic in QSM \cite{deistung2017overview}. For these reasons, and following \cite{bollmann2019deepqsm,chen2020qsmgan,jung2020exploring,yoon2018quantitative}, we use a high-quality COSMOS reconstruction from multiple phase as ground-truth. We minimize the $\ell_2$ loss function\footnote{We chose an $\ell_2$ norm for simplicity, and other choices are certainly possible, potentially providing further reconstruction improvements.} over the training samples via the following optimization formulation:
\begin{equation}
    \phi_{\hat{\theta}} = \arg\min_{\theta} \frac{1}{N} \sum_{j=1}^N \|x^j_c - \phi_\theta(y^j)\|^2_2,
\end{equation}
which is minimized with a stochastic first order method. We expand on the experimental details below.

\subsubsection{Deployment with Multiple Input Phase Measurements} Our proposed learned proximal network benefits from a clear separation between the utilization of the forward physical model and the learned proximal network parameters. This has two important advantages: (i) no ad-hoc rotation of the input image is required to align the main field ($B_0$), as this is instead managed by employing the operator $\Phi$ in the restoration; and (ii) the model learns a proximal function as a prior to susceptibility maps which is separated from the measurement model. More precisely, our learned proximal function has the same domain and co-domain, namely the space of susceptibility-map vectors, $P_{\hat{\theta}}:\mathcal{X}\to\mathcal{X}$. As a result, if at deployment (test) time the input measurements change, these can naturally be absorbed by modifying the function $f$ in Eq. \eqref{eq:regularized_LS}. In the important case where $L\geq 1$ measurements are available at arbitrary orientations, say $\{y_l = \Phi_l x\}_{l=1}^L$, a restored susceptibility map can still be computed with the obtained algorithm in Eq. \eqref{eq:learned_prox} by simply constructing the operator $\bar{\Phi} = [\Phi_1;\dots;\Phi_L] \in \mathbb{R}^{n L\times n}$ and concatenating the respective measurements, $\bar{y} = [y_1;\dots;y_L] \in \mathbb{R}^{n L} $, and performing the updates $\hat{x}_{i+1} = P_{\hat{\theta}} \left( \frac{\alpha}{L} \bar{\Phi} ^H \bar{y} + (I -\frac{\alpha}{L} \bar{\Phi}^H \bar{\Phi})\hat{x}_i\right).$
As we will demonstrate in the next section, the same LP-CNN model (trained with a single phase image) benefits from additional local phase inputs at test time, providing improved reconstructions without any re-training or tuning.

\section{Experiments and Results}
We trained and tested our LP-CNN method with our 8 subject dataset in a 4-fold cross validation manner, with each split containing 6 subjects for training and 2 subjects for testing. Our LP-CNN is implemented\footnote{Our code is released at \href{https://github.com/Sulam-Group}{https://github.com/Sulam-Group}.} in Pytorch and trained on a single NVIDIA GeForce RTX 2080 Ti GPU with mini-batch of size 2. We train our LP-CNN model with $k=3$ iterates for 100 epochs (96hrs) with dropout rate 0.5, using Adam optimizer \cite{kingma2014adam} with an initial learning rate $10^{-4}$ with weight decay of $5\times10^{-4}$, and we decrease the learning rate with 0.8 ratio every 25 epochs. We compare our results with traditional methods (TKD \cite{shmueli2009magnetic}, MEDI \cite{liu2011morphology} and STAR-QSM \cite{wei2015streaking} in STI Suite \cite{Li20135223SS}) as well as with the deep learning-based QSMnet \cite{yoon2018quantitative} across all splits.

\subsubsection{Patch Training} Training convolutional neural networks with 3D medical images often requires a prohibitively large amount of GPU memory. By taking advantage of the shift-invariance properties of CNN, a common workaround is to train these models in small patched data \cite{bollmann2019deepqsm,chen2020qsmgan,jung2020exploring,yoon2018quantitative}, we partition our data into $64\times64\times64$ 3D patches for training. However, such a modification must be done with caution, since naively employing a patched dipole kernel degrades performance by losing most of the high frequency information. In order to utilize the full dipole kernel during training with patched data, we modify the forward operator $\Phi$ by zero-padding, i.e. $\Phi' = CF^{-1}DFP = C\Phi P$,
where $P$ is a padding operation from $64^3$ to the original image size and $C$ is its (adjoint) cropping operator, $C = P^T$. See supplementary material for visualizing the difference between employing a patched dipole kernel and a full dipole kernel.

\subsubsection{Reconstruction Performance} 
\begin{figure}
\includegraphics[width=\textwidth]{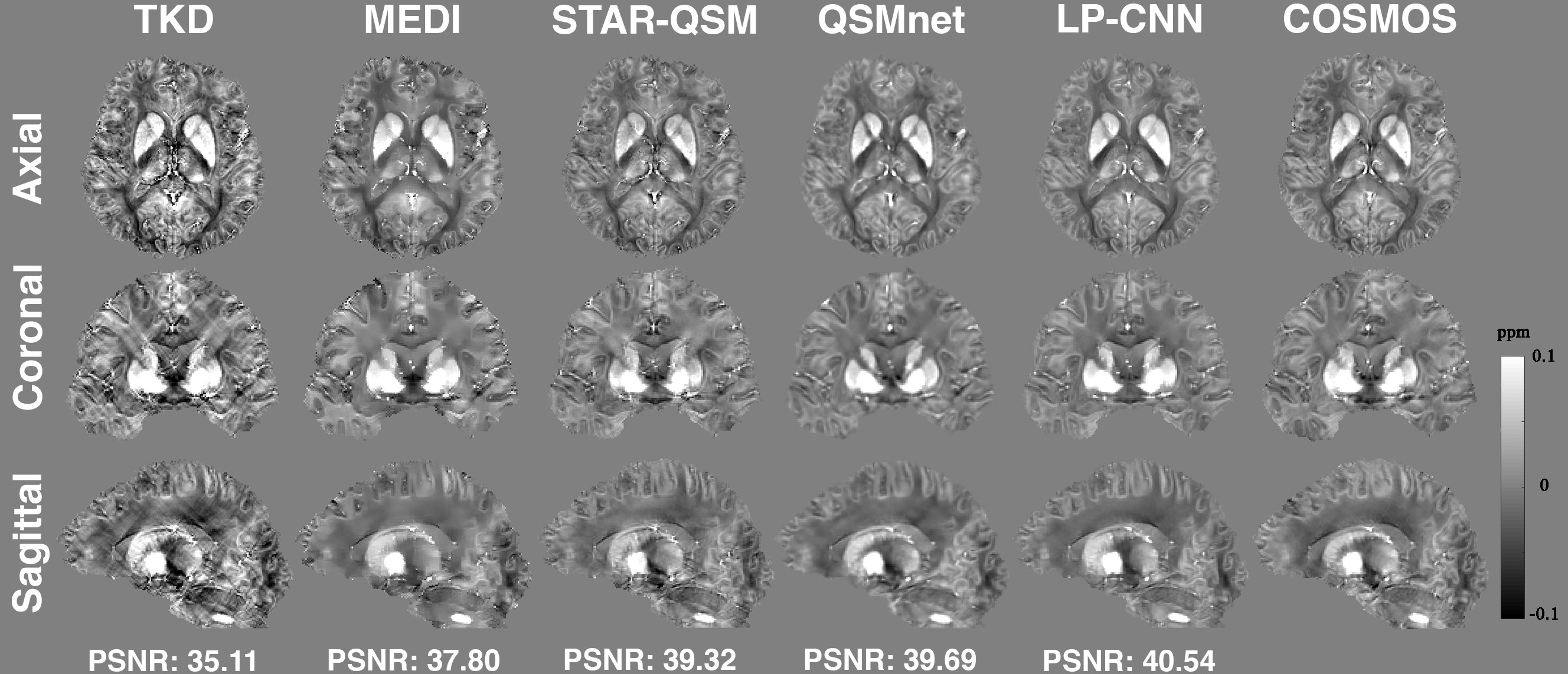}
\caption{Three plane views of susceptibility maps reconstructed using different methods. LP-CNN result shows high-quality COSMOS-like reconstruction performance. The TKD reconstruction suffers from streaking artifacts, while MEDI is over smoothing and STAR-QSM shows slight streaking effect.}
\label{fig2}
\end{figure}

Table~\ref{tab1} shows the results of our LP-CNN and other methods evaluated by quantitative performance metrics of Normalized Root Mean Square Error (NRMSE), Peak Signal-to-Noise Ratio (PSNR), High Frequency Error Norm (HFEN), and Structural Similarity Index (SSIM). We denote LP-CNN$^{L}$ as training with single input and testing with $L$ number multiple inputs. Fig.~\ref{fig2} shows the visualization results. The results of our LP-CNN showcases more high-frequency details and is the closest to the gold standard, COSMOS.

\begin{table}[t]
\centering
\caption{Quantitative performance metrics of QSM reconstruction estimated using the 4-fold cross validation set with different QSM methods. Upper part is single phase reconstruction and lower part is multiple phase reconstruction. LP-CNN shows comparable performance with QSMnet and outperforms TKD, MEDI and STAR-QSM.}
\label{tab1}
\begin{tabular}{|l||c|c|c|c|}
\hline
Methods&                        NRMSE($\%$)&  PSNR(dB)& HFEN($\%$)& SSIM\\
\hline
TKD&                $87.79\pm8.15$& $35.54\pm0.799$& $83.59\pm10.19$& $0.9710\pm0.0054$\\
MEDI&               $67.40\pm7.91$& $37.89\pm0.996$& $60.01\pm9.63$& $0.9823\pm0.0041$\\
STAR-QSM&          $63.70\pm5.68$& $38.32\pm0.758$& $62.06\pm6.19$& $0.9835\pm0.0032$\\
QSMnet&            $55.10\pm4.93$& $39.56\pm0.788$& $58.14\pm6.21$& $0.9857\pm0.0026$\\
LP-CNN&             $55.00\pm5.01$& $39.58\pm0.792$& $55.54\pm5.73$& $0.9865\pm0.0026$\\
\hline
\hline
LP-CNN$^{2}$&             $50.10\pm4.15$& $40.43\pm0.733$& $48.96\pm3.58$& $0.9889\pm0.0021$\\
LP-CNN$^{3}$&             $47.38\pm4.17$& $40.90\pm0.780$& $46.14\pm3.58$& $0.9903\pm0.0021$\\
\hline
\end{tabular}
\end{table}

\subsubsection{Multiple Input LP-CNN}
\begin{figure}
\includegraphics[width=\textwidth]{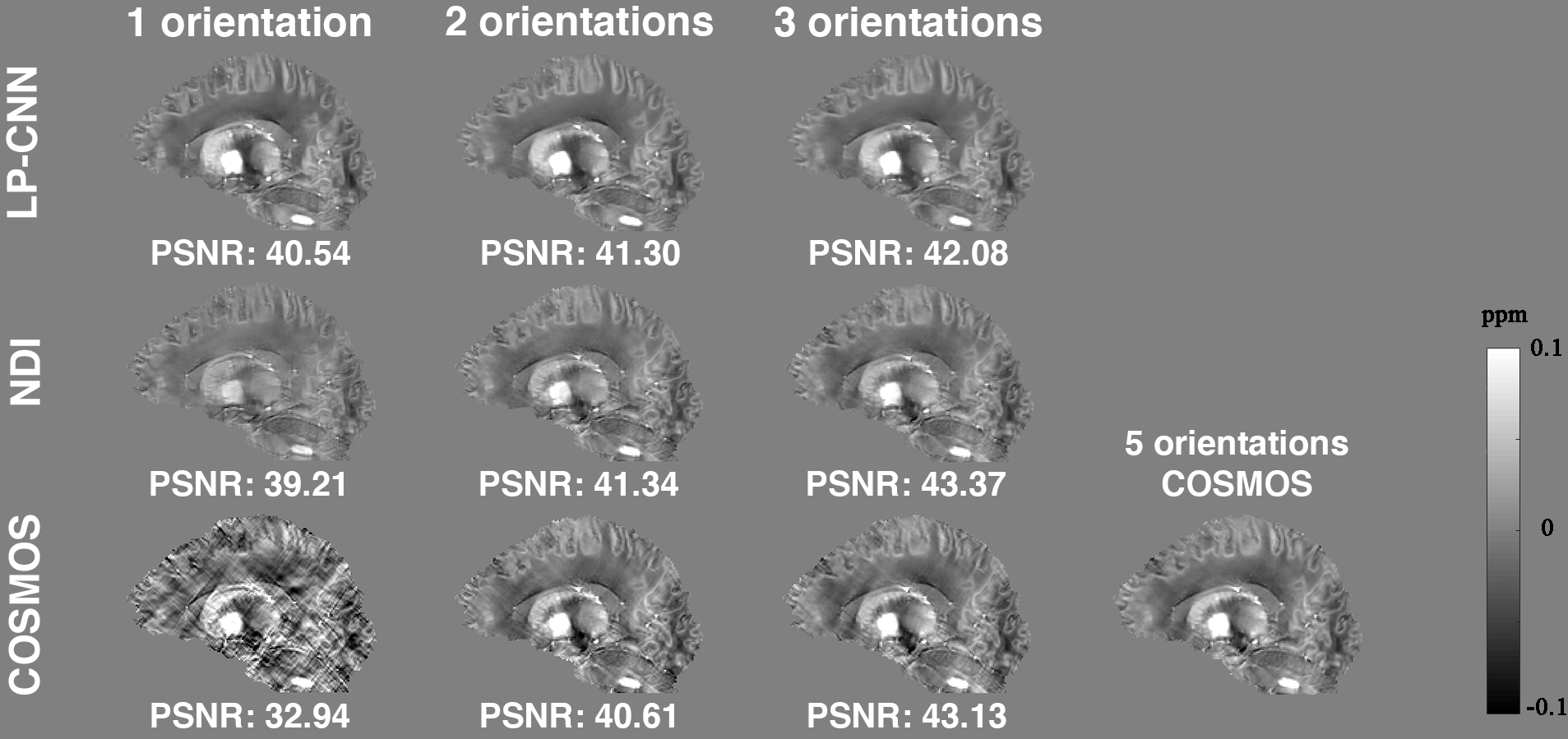}
\caption{Comparison between LP-CNN, NDI and COSMOS for multiple numbers of input phase measurement reconstructions. LP-CNN shows high-quality reconstruction on single input and refine the reconstruction with multiple inputs.} 
\label{fig3}
\end{figure}

We finally demonstrate the flexibility of our LP-CNN by comparing the performance of our obtained model (trained with single local phase measurements) when deployed with increasing number of input measurements, and we compare with NDI \cite{NDI} and COSMOS using a different number of inputs in Fig.~\ref{fig3}. Though our LP-CNN was only trained with single local phase input, it allows to incorporate phase measurements at multiple orientations that improve QSM reconstruction, while none of the previous deep learning-based QSM methods \cite{jung2020overview} can handle a different number of input phase measurements.

\section{Discussion and Conclusion}
In this work, we proposed a LP-CNN method to resolve the ill-posed dipole inversion problem in QSM. Our approach naturally incorporates the forward operator with arbitrarily oriented dipole and enhances reconstruction accuracy for single orientation QSM giving COSMOS like solution. More importantly, the learned parameters are separated from the measurement operator, providing greater flexibility. In particular, we demonstrated how reconstruction performance increases with the number of input local phase images at test time; even while the LP-CNN network was trained with a single phase measurement. Furthermore, in order to address tissue magnetic susceptibility anisotropy \cite{deistung2017overview}, similar tools could be also extended to more complex, higher order, susceptibility models such as in Susceptibility Tensor Imaging (STI) \cite{li2017susceptibility,liu2010susceptibility}.

\subsubsection{Acknowledgements.} The authors would like to thank Mr. Joseph Gillen, Ms. Terri Brawner, Ms. Kathleen Kahl and Ms. Ivana Kusevic for their assistance with data acquisition. This work was partly supported by NCRR and NIBIB (P41 EB015909).

%
%
%
\bibliographystyle{splncs04}
\bibliography{reference}

\begin{thebibliography}{10}
\providecommand{\url}[1]{\texttt{#1}}
\providecommand{\urlprefix}{URL }
\providecommand{\doi}[1]{https://doi.org/#1}

\bibitem{abdul2005fast}
Abdul-Rahman, H., et~al.: Fast three-dimensional phase-unwrapping algorithm
  based on sorting by reliability following a non-continuous path. In: Optical
  Measurement Systems for Industrial Inspection IV. vol.~5856, pp. 32--40.
  International Society for Optics and Photonics (2005)

\bibitem{acosta2013vivo}
Acosta-Cabronero, J., et~al.: In vivo quantitative susceptibility mapping (qsm)
  in alzheimer's disease. PloS one  \textbf{8}(11) (2013)

\bibitem{adler2018learned}
Adler, J., {\"O}ktem, O.: Learned primal-dual reconstruction. IEEE transactions
  on medical imaging  \textbf{37}(6),  1322--1332 (2018)

\bibitem{ayton2017cerebral}
Ayton, S., et~al.: Cerebral quantitative susceptibility mapping predicts
  amyloid-$\beta$-related cognitive decline. Brain  \textbf{140}(8),
  2112--2119 (2017)

\bibitem{bollmann2019deepqsm}
Bollmann, S., et~al.: Deepqsm-using deep learning to solve the dipole inversion
  for quantitative susceptibility mapping. NeuroImage  \textbf{195},  373--383
  (2019)

\bibitem{bruckstein2009sparse}
Bruckstein, A.M., et~al.: From sparse solutions of systems of equations to
  sparse modeling of signals and images. SIAM review  \textbf{51}(1),  34--81
  (2009)

\bibitem{chen2014quantitative}
Chen, W., et~al.: Quantitative susceptibility mapping of multiple sclerosis
  lesions at various ages. Radiology  \textbf{271}(1),  183--192 (2014)

\bibitem{chen2020qsmgan}
Chen, Y., et~al.: Qsmgan: Improved quantitative susceptibility mapping using 3d
  generative adversarial networks with increased receptive field. NeuroImage
  \textbf{207},  116389 (2020)

\bibitem{de2008quantitative}
De~Rochefort, L., et~al.: Quantitative mr susceptibility mapping using
  piece-wise constant regularized inversion of the magnetic field. Magnetic
  Resonance in Medicine: An Official Journal of the International Society for
  Magnetic Resonance in Medicine  \textbf{60}(4),  1003--1009 (2008)

\bibitem{deistung2017overview}
Deistung, A., et~al.: Overview of quantitative susceptibility mapping. NMR in
  Biomedicine  \textbf{30}(4),  e3569 (2017)

\bibitem{goodfellow2014generative}
Goodfellow, I., et~al.: Generative adversarial nets. In: Advances in neural
  information processing systems. pp. 2672--2680 (2014)

\bibitem{hammernik2018learning}
Hammernik, K., et~al.: Learning a variational network for reconstruction of
  accelerated mri data. Magnetic resonance in medicine  \textbf{79}(6),
  3055--3071 (2018)

\bibitem{jenkinson2001global}
Jenkinson, M., Smith, S.: A global optimisation method for robust affine
  registration of brain images. Medical image analysis  \textbf{5}(2),
  143--156 (2001)

\bibitem{jung2020overview}
Jung, W., Bollmann, S., Lee, J.: of quantitative susceptibility mapping using
  deep learning: Current status, challenges and opportunities. NMR in
  Biomedicine p. e4292 (2020)

\bibitem{jung2020exploring}
Jung, W., et~al.: Exploring linearity of deep neural network trained qsm:
  Qsmnet+. NeuroImage p. 116619 (2020)

\bibitem{pvn}
Kames, C., et~al.: Proximal variational networks: generalizable deep networks
  for solving the dipole-inversion problem. 5th International QSM Workshop
  (2019)

\bibitem{proxvnet}
Kames, C., et~al.: Proxvnet: A proximal gradient descent-based deep learning
  model for dipole inversion in susceptibility mapping. International Society
  for Magnetic Resonance in Medicine  (2020)

\bibitem{kingma2014adam}
Kingma, D.P., Ba, J.: Adam: A method for stochastic optimization. arXiv
  preprint arXiv:1412.6980  (2014)

\bibitem{langkammer2013quantitative}
Langkammer, C., et~al.: Quantitative susceptibility mapping in multiple
  sclerosis. Radiology  \textbf{267}(2),  551--559 (2013)

\bibitem{langkammer2018quantitative}
Langkammer, C., et~al.: Quantitative susceptibility mapping: report from the
  2016 reconstruction challenge. Magnetic resonance in medicine
  \textbf{79}(3),  1661--1673 (2018)

\bibitem{Li20135223SS}
Li, W., Wu, B., Liu, C.: 5223 sti suite : a software package for quantitative
  susceptibility imaging (2013)

\bibitem{li2017susceptibility}
Li, W., et~al.: Susceptibility tensor imaging (sti) of the brain. NMR in
  Biomedicine  \textbf{30}(4),  e3540 (2017)

\bibitem{li2016magnetic}
Li, X., et~al.: Magnetic susceptibility contrast variations in multiple
  sclerosis lesions. Journal of magnetic resonance imaging  \textbf{43}(2),
  463--473 (2016)

\bibitem{liu2010susceptibility}
Liu, C.: Susceptibility tensor imaging. Magnetic Resonance in Medicine: An
  Official Journal of the International Society for Magnetic Resonance in
  Medicine  \textbf{63}(6),  1471--1477 (2010)

\bibitem{liu2015quantitative}
Liu, C., et~al.: Quantitative susceptibility mapping: contrast mechanisms and
  clinical applications. Tomography  \textbf{1}(1), ~3 (2015)

\bibitem{liu2015susceptibility}
Liu, C., et~al.: Susceptibility-weighted imaging and quantitative
  susceptibility mapping in the brain. Journal of magnetic resonance imaging
  \textbf{42}(1),  23--41 (2015)

\bibitem{liu2009calculation}
Liu, T., et~al.: Calculation of susceptibility through multiple orientation
  sampling (cosmos): a method for conditioning the inverse problem from
  measured magnetic field map to susceptibility source image in mri. Magnetic
  Resonance in Medicine: An Official Journal of the International Society for
  Magnetic Resonance in Medicine  \textbf{61}(1),  196--204 (2009)

\bibitem{liu2011morphology}
Liu, T., et~al.: Morphology enabled dipole inversion (medi) from a single-angle
  acquisition: comparison with cosmos in human brain imaging. Magnetic
  resonance in medicine  \textbf{66}(3),  777--783 (2011)

\bibitem{liu2013nonlinear}
Liu, T., et~al.: Nonlinear formulation of the magnetic field to source
  relationship for robust quantitative susceptibility mapping. Magnetic
  resonance in medicine  \textbf{69}(2),  467--476 (2013)

\bibitem{mardani2018deep}
Mardani, M., et~al.: Deep generative adversarial neural networks for
  compressive sensing mri. IEEE transactions on medical imaging
  \textbf{38}(1),  167--179 (2018)

\bibitem{osher2005iterative}
Osher, S., et~al.: An iterative regularization method for total variation-based
  image restoration. Multiscale Modeling \& Simulation  \textbf{4}(2),
  460--489 (2005)

\bibitem{ozbay2017comprehensive}
{\"O}zbay, P.S., et~al.: A comprehensive numerical analysis of background phase
  correction with v-sharp. NMR in Biomedicine  \textbf{30}(4),  e3550 (2017)

\bibitem{parikh2014proximal}
Parikh, N., et~al.: Proximal algorithms. Foundations and
  Trends{\textregistered} in Optimization  \textbf{1}(3),  127--239 (2014)

\bibitem{NDI}
Polak, D., et~al.: {Nonlinear dipole inversion (NDI) enables robust
  quantitative susceptibility mapping (QSM)}. NMR in Biomedicine  (2020).
  \doi{10.1002/nbm.4271}

\bibitem{ronneberger2015u}
Ronneberger, O., et~al.: U-net: Convolutional networks for biomedical image
  segmentation. In: International Conference on Medical image computing and
  computer-assisted intervention. pp. 234--241. Springer (2015)

\bibitem{shmueli2009magnetic}
Shmueli, K., et~al.: Magnetic susceptibility mapping of brain tissue in vivo
  using mri phase data. Magnetic Resonance in Medicine: An Official Journal of
  the International Society for Magnetic Resonance in Medicine  \textbf{62}(6),
   1510--1522 (2009)

\bibitem{smith2002fast}
Smith, S.M.: Fast robust automated brain extraction. Human brain mapping
  \textbf{17}(3),  143--155 (2002)

\bibitem{sulam2019multi}
Sulam, J., et~al.: On multi-layer basis pursuit, efficient algorithms and
  convolutional neural networks. IEEE transactions on pattern analysis and
  machine intelligence  (2019)

\bibitem{van2016colocalization}
Van~Bergen, J., et~al.: Colocalization of cerebral iron with amyloid beta in
  mild cognitive impairment. Scientific reports  \textbf{6}(1), ~1--9 (2016)

\bibitem{wang2015quantitative}
Wang, Y., Liu, T.: Quantitative susceptibility mapping (qsm): decoding mri data
  for a tissue magnetic biomarker. Magnetic resonance in medicine
  \textbf{73}(1),  82--101 (2015)

\bibitem{wei2015streaking}
Wei, H., et~al.: Streaking artifact reduction for quantitative susceptibility
  mapping of sources with large dynamic range. NMR in Biomedicine
  \textbf{28}(10),  1294--1303 (2015)

\bibitem{yoon2018quantitative}
Yoon, J., et~al.: Quantitative susceptibility mapping using deep neural
  network: Qsmnet. NeuroImage  \textbf{179},  199--206 (2018)

\bibitem{zagoruyko2016wide}
Zagoruyko, S., Komodakis, N.: Wide residual networks. arXiv preprint
  arXiv:1605.07146  (2016)

\end{thebibliography}

\end{document}